\documentclass[twocolumn,showpacs,pra]{revtex4}
\usepackage{amsmath,amssymb}
\usepackage{graphicx}

\begin{document}

\title{Cold atom dynamics in non-Abelian gauge fields}
\author{A. Jacob$^1$, P. \"Ohberg$^2$, G. Juzeli\=unas$^3$ and L. Santos$^1$} 
\affiliation{
\mbox{$^1$Institut f\"ur Theoretische Physik , Leibniz Universit\"at
Hannover, Appelstr. 2, D-30167, Hannover, Germany}\\
\mbox{$^2$ SUPA, Department of Physics, Heriot-Watt University, Edinburgh, UK}\\
\mbox{$^3$Institute of Theoretical Physics and Astronomy of Vilnius 
University, A. Go\v{s}tauto 12, 01108 Vilnius, Lithuania}\\
}

\begin{abstract}
The dynamics of ultracold neutral atoms subject to a 
non-Abelian gauge field is investigated. In particular we analyze in detail 
a simple experimental scheme to achieve a constant, but non-Abelian gauge field, and 
discuss in the frame of this gauge field the non-Abelian Aharanov-Bohm effect. 
In the last part of this paper, we discuss intrinsic non-Abelian effects in the dynamics 
of cold atomic wavepackets.  
\end{abstract}

\maketitle

\section{Introduction}
\label{sec:1}

Gauge potentials, and gauge theories in general, are crucial for 
the understanding of the fundamental forces between 
subatomic particles. The simplest example of gauge potentials is the 
vector potential in the theory of electromagnetism \cite{Jackson}, which is an example 
of an Abelian gauge field. Non-Abelian situations, where the gauge potential 
is a matrix whose vector components do not commute, are surprisingly scarce 
in Nature. Candidates so far have mainly been restricted to molecular systems \cite{mead1992}
which are largely approachable only through spectroscopic means. Other systems 
are liquid crystals which show the required non-Abelian symmetries
\cite{lavrentovich1986,chernodub2006}. 

An elegant derivation and description of 
the emergence of non-Abelian gauge potentials 
has been presented by Wilczek and Zee \cite{wilczek1984}. It was 
shown by these authors that in the presence of a general adiabatic motion of 
a quantum system with degenerate states, gauge potentials will appear 
which are traditionally only encountered in high energy physics to 
describe the interactions between elementary particles. 
Ultracold atomic clouds are particularly promising candidates 
for realising such scenarios, since the  
access to physical parameters is, from an experimental point of view, 
unprecedented. Extending the ideas of Wilczek and Zee, it was 
recently proposed that properly tailored laser beams coupled to degenerated 
internal electronic states can be employed to induce non-Abelian 
gauge fields in cold-atom experiments \cite{ruseckas2005}.
Alternatively, a non-Abelian gauge potential can be constructed in an optical lattice 
using laser assisted state sensitive tunneling \cite{osterloh2005}.
With the implementation of these proposals, ultracold atoms would offer 
a unique testbed for the analysis of non-trivial non-Abelian effects on 
the quantum dynamics of multicomponent wavepackets.

In this paper we investigate the wave packet dynamics of a cloud 
of ultracold atoms in the presence of a non-Abelian gauge potential. 
In Sec.~\ref{sec:2} we discuss how this undoubtedly rather exotic scenario 
can be envisaged in a sample of cold atoms where the internal electronic 
energy levels are addressed by laser fields with a nontrivial spatial 
phase and intensity distribution. This setup opens up a 
number of new scenarios for ultracold gases, 
allowing for the study of {\em non-Abelian atom optics}, which  
naturally ties together optical and magnetic effects. Remarkably, as shown 
in Sec.~\ref{sec:2}, even very simple laser arrangements may induce 
non-trivial cold-atom dynamics. As a first example of this 
non-trivial dynamics, we discuss in Sec.~\ref{sec:3} a possible optical tweezer
experiment including a non-Abelian flux, for which the population transfer 
between internal states crucially depends on the path taken 
(non-Abelian Aharonov-Bohm effect). This effect resembles indeed 
what one would expect from scattering protons onto a non-Abelian flux 
line where the proton can be transfered into a neutron \cite{horvathy1986}.
The tweezer experiment discussed in Sec.~\ref{sec:3} just involves 
the internal-state dynamics, without exploring the rich dynamics 
resulting from the interplay between external and internal degrees of freedom 
in non-Abelian gauge fields. Sec.~\ref{sec:4} is devoted to the 
analysis of this interplay. In particular, we show that the dynamics of 
cold-atom wavepackets can be significantly affected by intrinsically non-Abelian effects, which 
are crucially dependent on the initial momentum distribution of the
wavepacket. We consider in particular the relevant examples of wavepacket 
propagation and wavepacket reflection at an atomic mirror. Finally, we
conclude in Sec.~\ref{sec:5}.

\begin{figure}[h]
\begin{center}
\includegraphics[width=8cm]{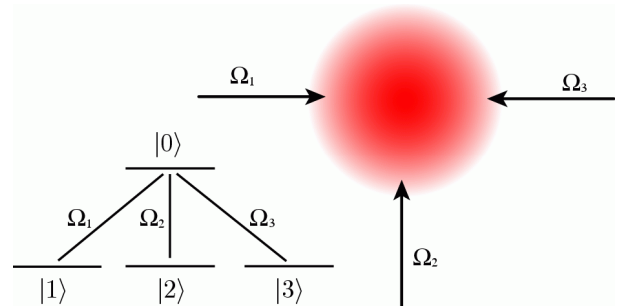}
\end{center}
\caption{The tripod coupling scheme forms two degenerate dark states with a
non-adiabatic coupling. The three laser beams $\Omega_i$ $i=1,2,3$ are 
arranged as two counter propagating beams $(\Omega_1$ and $\Omega_2$) 
and one beam $(\Omega_3)$ (of double intensity) in the perpendicular direction.}
\label{fig:1}
\end{figure}

\section{Laser-induced non Abelian Gauge fields}
\label{sec:2}


In a recent paper \cite{ruseckas2005} it was shown that a non-Abelian gauge 
potential can be constructed in the presence of nontrivial light fields 
coupled to degenerate electronic states of cold atoms. For this we consider
atoms with multiple internal states, see Fig.\ (1).  
For a fixed position $\mathbf{r}$ the internal
Hamiltonian $\hat{H}_{0}(\mathbf{r})$ including the laser interaction can be
diagonalized to give a set of $4$ dressed states $\left| \chi_n(\mathbf{r})
\right\rangle $ with eigenvalues $\varepsilon _{n}(\mathbf{r})$, where
$n=1,2,3,4$. The full quantum state of the atom describing both internal
and motional degrees of freedom can then be expanded in terms of the dressed
states according to $|\Phi\rangle =\sum_{n=1}^{4}\Psi _{n}(\mathbf{r}) \left|
\chi_n(\mathbf{r})\right\rangle$.  If there are two degenerate dressed states and we can neglect the
transitions to the other states we obtain a coupled two level system of the form
\begin{equation}
i\hbar\frac{\partial}{\partial t}\tilde\Psi=\left[ \frac{1}{2m}
(-i\hbar\nabla -\mathbf{A})^2 + V +\Phi \right]\tilde\Psi,
\label{eq:SE-reduced}
\end{equation}
where the $2\times2$ potentials are given by
\begin{align}
V_{n,m} &= \varepsilon _{n}(\mathbf{r})\, \delta_{n,m}
+\langle \chi_n(\mathbf{r})|V(\mathbf{r})|\chi_m(\mathbf{r})\rangle ,
\label{eq:V}\\
\mathbf{A}_{n,m} &= i\hbar \langle \chi_n(\mathbf{r})|
\nabla\chi_m(\mathbf{r}) \rangle . \label{eq:A} \\
\Phi _{n,m} &=\frac{\hbar ^2}{2m}\left( \langle\nabla\chi _{n}|\nabla\chi_{m}\rangle
+\sum_{k=1}^{2}\langle\chi _{n}|\nabla\chi _{k}\rangle
\langle\chi _{k}|\nabla\chi _{m}\rangle \right)  \label{eq:fi1} 
\end{align}
The  reduced 
$2\times 2$ matrix $\mathbf{A}$ is sometimes referred to as
the Berry connection and is related to a curvature (an effective
``magnetic'' field) $\mathbf{B}$ as
\begin{eqnarray}
B_i = \frac{1}{2}\epsilon_{ikl} \, F_{kl},\quad
F_{kl} = \partial_k A_l-\partial_l A_k -\frac{i}{\hbar}[A_k,A_l].\label{eq:B}
\end{eqnarray}
Note that the term $\frac{1}{2}\varepsilon_{ikl}[A_k,A_l] = 
(\mathbf{A}\times \mathbf{A})_i$ does not vanish in general,
since the vector components of $\mathbf{A}$ do not necessarily commute. 
In fact this
term reflects the non-Abelian character of the gauge potentials.

To construct a scheme of laser-atom interactions that leads to a
$\sf{U}(2)$ gauge potential we need
degenerate (or nearly degenerate) dressed states.  Such a condition is
fulfilled e.g.\  for the tripod system shown in Fig.\ \ref{fig:1}.
A truly non-Abelian situation
emerges if the matrices $\{A_x,A_y,A_z,\Phi\}$ do not commute. 
For this it is necessary that the off-diagonal element $i\hbar \langle\chi_1(\mathbf{r})|\nabla
\chi_2(\mathbf{r})\rangle$ is non-zero.
The Hamiltonian of the tripod system reads in interaction representation as 
\cite{ruseckas2005,unanyan98,unanyan99}
\begin{equation}
\hat{H}_0=-\hbar\Bigl(\Omega_1|0\rangle\langle 1|+\Omega_2|0\rangle\langle 2|
+\Omega_3|0\rangle\langle 3|\Bigr)+H.c.,
\end{equation}
Parameterizing the Rabi-frequencies $\Omega_\mu$ with angle and phase variables
according to
$\Omega_1  = \Omega\, \sin\theta\, \cos\phi\, \mathrm{e}^{iS_1}, 
\Omega_2  = \Omega\, \sin\theta\, \sin\phi\,  \mathrm{e}^{iS_2},
\Omega_3  = \Omega\, \cos\theta\, \mathrm{e}^{iS_3},$
where $\Omega =\sqrt{|\Omega_1|^2+|\Omega_2|^2+|\Omega_3|^2}$, the adiabatic
dark states read
\begin{eqnarray}
|D_1\rangle & = &\sin\phi \mathrm{e}^{iS_{31}}|1\rangle
-\cos\phi \mathrm{e}^{iS_{32}}|2\rangle,
\label{eq:D1} \\
|D_2\rangle & = &\cos\theta \cos\phi \mathrm{e}^{iS_{31}}|1\rangle
+\cos\theta\sin\phi \mathrm{e}^{iS_{32}}|2\rangle \nonumber\\
&&  -\sin\theta |3\rangle ,
\label{eq:D2}
\end{eqnarray}
with $S_{ij}=S_i-S_j$. 

The gauge potential depends on the gradient of the dark states:
\begin{eqnarray}
\mathbf{A}_{11} &=& \hbar\left(\cos^2\phi\nabla S_{23}
+ \sin^2\phi\nabla S_{13}\right)\, ,\nonumber \\
\mathbf{A}_{12} &=& \hbar\cos\theta\left(\frac{1}{2}\sin(2\phi)
\nabla S_{12}-i\nabla\phi\right)\, , \label{eq:A-special} \\
\mathbf{A}_{22} &=&\hbar\cos^2\theta\left(\cos^2\phi
\nabla S_{13}+\sin^2\phi\nabla S_{23}\right),\nonumber
\end{eqnarray}
and
\begin{align}
\Phi_{11} & = \frac{\hbar^2}{2m}\sin^2\theta\left(\frac{1}{4}
\sin^2(2\phi)(\nabla S_{12})^2+(\nabla\phi)^2\right),
\nonumber \\
\Phi_{12} & = \frac{\hbar^2}{2m}\sin\theta
\left(\frac{1}{2}\sin(2\phi)\nabla S_{12}
-i\nabla\phi\right)\\
 & \left(\frac{1}{2}\sin(2\theta)(\cos^2\phi\nabla S_{13}
+\sin^2\phi\nabla S_{23})-i\nabla\theta\right), \nonumber\\
\Phi_{22} & = \frac{\hbar^2}{2m}\biggl(\frac{1}{4}\sin^2(2\theta)\left(
\cos^2\phi\nabla S_{13}+\sin^2\phi\nabla S_{23}\right)^2 \nonumber \\
& +(\nabla\theta)^2\biggr).\nonumber
\end{align}
This provides a remarkable versatility. Recent advances in shaping 
both the phase and the intensity of light beams makes it possible 
to choose practically any shape of the gauge potential provided the 
corresponding light field obeys Maxwell's equations. This is 
certainly the case in a two-dimensional geometry, but also in three 
dimensions light beams can be tailored \cite{mcgloin03,whyte2005}. In the Abelian case a nonzero 
effective magnetic field is obtained if there is a relative angular 
momentum between the two light beams and the intensity ratio is 
spatially dependent \cite{juzeliunas2004,juzeliunas2005,juzeliunas2006}. 

Surprisingly, the generation of a non-Abelian gauge field 
does not require any elaborate 
shaping of the three laser beams employed. This is indeed the case if we
choose the configuration shown in Fig.~\ref{fig:1}. Three plane-wave laser
beams are used. Two lasers of equal intensity are counter-propagating in the 
$x$-direction with wave vector $\kappa$ while the
third one (of double intensity) propagates in the y-direction also with a wave
vector $\kappa$. With this arrangement, $\phi=\pi/4$, and $\theta=\pi/4$ in the
expressions above. The resulting vector potential is of the form:
\begin{equation}
\mathbf{A}  =  \hbar\kappa\left(\begin{array}{cc}
-\mathbf{e}_{y} & \mathbf{e}_{x}/\sqrt{2}\\
\mathbf{e}_{x}/\sqrt{2} & -\mathbf{e}_{y}/2\end{array}\right),
\label{eq:A-Matrix}
\end{equation}
whereas
\begin{equation}
V+\Phi  =  \left(\begin{array}{cc}
V_{1}+\frac{\hbar^{2}\kappa^{2}}{4m}& 0\\
0 & (V_{1}+V_{3})/2+ \frac{\hbar^{2}\kappa^{2}}{8m}
\end{array}\right),\label{eq:V-Matrix}
\end{equation}
By choosing the laser detuning such that 
$V_{3}-V_{1}=\hbar^{2}\kappa^{2}/4m$ we obtain a scalar potential 
proportional to the unit matrix, $V+\Phi=V_{1}\mathrm{I}$.
Therefore the scalar potential can be safely neglected as far as the wavepacket dynamics is concerned.

\section{Non-Abelian Aharonov-Bohm effect} 
\label{sec:3}

\begin{figure}
\includegraphics[width=8 cm]{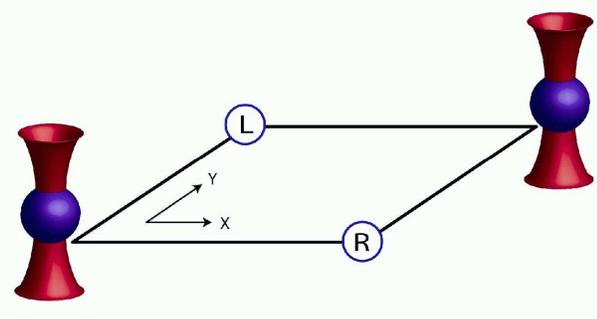}
\caption{(Color online). The envisaged experiment. 
An optical tweezer moves the cloud of atoms along the left 
(L) path or the right (R) path. The final dark state 
population will depend on which path was taken.}
\label{fig:2}
\end{figure}

\begin{figure}
\includegraphics[width=6 cm]{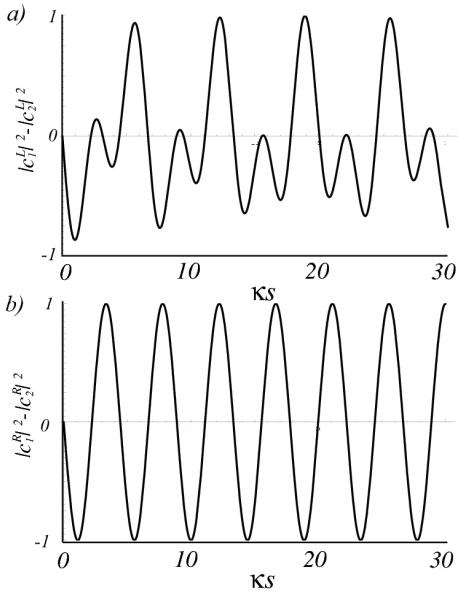}
\caption{The difference in the populations in the two 
dark states depends on which path is taken. 
Fig. (a) shows the total difference $|c_1^L|^2-|c_2^L|^2$ as a function of the
path length $\kappa\sigma$, whereas in Fig. (b)
we depict $|c_1^R|^2- |c_2^R|^2$. We assume as initial condition 
$\eta=\pi/4$, and $\varphi=\pi/2$.}
\label{fig:3}
\end{figure}

\begin{figure}
\includegraphics[width=8 cm]{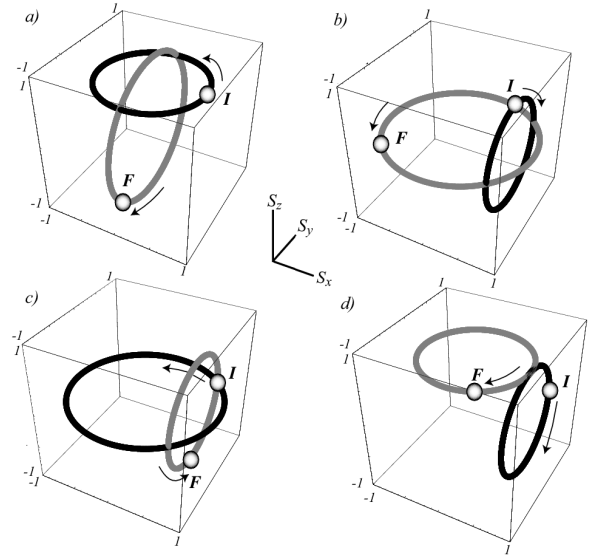}
\caption{The pseudo spin trajectories depend on the initial 
state and which path is taken: a) Left path with $\eta=\pi/8, 
\varphi=0$, b) Right path with $\eta=\pi/8, \varphi=0$, c) 
Left path with $\eta=\pi/4, \varphi=\pi/4$, d) 
Right path with $\eta=\pi/4, \varphi=\pi/4$. The spheres in each 
figure indicate the initial state (I) and the final state (F). 
The black circle is always the path first embarked on. In all cases we have chosen $\kappa s=34.5$. This will cause the spin vector to traverse the circular paths several times in each plane.}
\label{fig:4}
\end{figure}

In Ref.~\cite{osterloh2005} it was proposed that non-Abelian gauge fields 
created in lattices can be employed to construct non-Abelian atom
interferometers. However, the read-out of any non-Abelian atom interferometer 
may be crucially handicapped by the non-trivial interplay 
between external and internal degrees of freedom in the wavepacket dynamics 
of atoms in non-Abelian gauge fields (see Sec.~\ref{sec:4}). However, 
this coupling between external and internal dynamics may be prevented 
by considering atoms trapped in mobile optical tweezers. If the tweezer potential 
is strong enough, the system may be investigated in the so-called single-mode 
approximation, in which both components share exactly the same center-of-mass 
wavepacket. As a consequence, the non-Abelian gauge field will just affect the 
internal dynamics of the atoms. In the following we envisage 
an experiment in which a cloud of ultra cold atoms is trapped by an optical 
tweezer under the conditions discussed above. When moving in the xy plane the
atoms experience the gauge potential given by Eq.~(\ref{eq:A-Matrix}). 
We consider the case where the atoms are moved in the x and y direction 
(Fig.~\ref{fig:2}) along two different paths: (clock-wise, $L$) from $(0,0)$ to $(0,s)$ and then
from $(0,s)$ to $(s,s)$; (anti clock-wise, $R$) from $(0,0)$ to $(s,0)$ and then
from $(s,0)$ to $(s,s)$. The initial state of the atom is assumed 
to be a linear superposition of both dark states:
\begin{equation}
|\Psi(0)\rangle=\cos(\eta)|D_1\rangle+e^{i\varphi}\sin(\eta)|D_2\rangle
\end{equation}
where $\eta$ is the mixing angle, and $\varphi$ is a relative phase. 
The dynamics of the two level system obviously depends on 
the initial state, but more importantly, the final populations 
of the two dark states depend on which path is taken. 
After performing the clock-wise path the atoms are in the state
\begin{equation}
|\Psi_L\rangle=e^{i \hat A_x s/\hbar}e^{i \hat A_y s/\hbar}|
\Psi(0)\rangle=c_1^L |D_1\rangle+c_2^L |D_2\rangle
\end{equation}
whereas after performing the anti clock-wise path we have:
\begin{equation}
|\Psi_R\rangle=e^{i \hat A_y s/\hbar}e^{i \hat A_x s/\hbar}
|\Psi(0)\rangle=c_1^R |D_1\rangle+c_2^R |D_2\rangle.
\end{equation}  
Using the vector potential given by Eq.~(\ref{eq:A-Matrix}) 
a straight forward calculation yields
\begin{widetext}
\begin{eqnarray}
c_1^L&=&e^{-i\kappa s}
\cos(\frac{\kappa s}{\sqrt{2}})\cos(\eta)\!+
i e^{i(\varphi-\kappa s/2)}\sin(\frac{\kappa s}{\sqrt{2}})\sin(\eta)\\
c_2^L&=& ie^{-i\kappa s}
\sin(\frac{\kappa s}{\sqrt{2}})\cos(\eta)
+e^{i(\varphi-\kappa s/2)}\cos(\frac{\kappa s}{\sqrt{2}})\sin(\eta)\\
c_1^R&=&e^{-i\kappa s}
(\cos(\frac{\kappa s}{\sqrt{2}})\cos(\eta)+i 
e^{i\varphi}\sin(\frac{\kappa s}{\sqrt{2}})\sin(\eta))\\
c_2^R&=& e^{-i\kappa s/2}
(i\sin(\frac{\kappa s}{\sqrt{2}})\cos(\eta)+
e^{i\varphi}\cos(\frac{\kappa s}{\sqrt{2}})\sin(\eta)).
\end{eqnarray}
\end{widetext}
Fig.~\ref{fig:3} shows the final population difference 
between the two dark states for both paths as a function of the 
path length $\kappa s$. It becomes clear that the outcome of choosing 
the $L$ or $R$ path can be very different. We stress that this effect 
is not directly linked to the appearance of off-diagonal terms 
in the corresponding matrices of the vector potential, but rather 
it is inherently due to the non-Abelian character of the matrices 
$\hat A_x$ and $\hat A_y$. This effect is remarkably 
similar to the scattering of protons onto a non-Abelian flux line, 
where a conversion of the proton into a neutron is anticipated 
\cite{horvathy1986}. A more complete picture is obtained by 
defining the pseudo spin ${\bf S}(c_1^{L,R},c_2^{L,R})$ as 
\begin{eqnarray}
S_x&=& \frac{1}{2i}(c_1 {c_2}^*-{c_1}^* c_2)\\
S_y&=& \frac{1}{2}(c_1 {c_2}^*+{c_1}^* c_2)\\
S_z&=& \frac{1}{2}(|c_1|^2-|c_2|^2).
\end{eqnarray}
With the pseudo spin representation we can follow the rotation of the 
spin vector as a function of position along the different paths. 
This is shown in Fig.~\ref{fig:4} where the spin vector is 
seen to follow circular paths whose orientation changes when 
the direction of the atoms in real space changes. The role 
of the initial state is now immediately clear. Only a 
superposition between $|D_1\rangle$ and $|D_2\rangle$ 
will result in a different final state of ${\bf S}$ as 
a function of taking either the $L$ or $R$ path. 
Note in contrast to the 
previously considered laser-driven population transfer for 
tripod atoms \cite{unanyan98,unanyan99} here the non-Abelian dynamics 
is due to the time-dependence of the phases of light fields "seen" 
by moving atoms rather than due to the time-dependence of the 
intensities of laser pulses.

\section{Wavepackets in free space}
\label{sec:4}

The non-Abelian Aharonov-Bohm effect is a striking example where the 
internal dynamics of a two-level system is highly nontrivial. 
A question which is not often addressed in the context of non-Abelian systems 
is the dynamics of a wave packet. This situation is clearly more complex 
compared to the previous non-Abelian Aharonov-Bohm scenario where an adiabatic motion 
with respect to center of mass excitations and shape oscillations was assumed. 
We now have to fully take into account the coupled internal and external degrees of freedom. 

In the following we discuss the evolution of a cold atomic wavepacket in the 
presence of a non-Abelian gauge field $\hat A=\{\hat A_x,\hat A_y,0 \}$. 
We consider that the atomic gas is sufficiently dilute, and hence in this paper 
we neglect the effects of the interatomic interactions. 
We restrict ourselves to the case in which both matrices 
$\hat A_x$ and $\hat A_y$ are space-independent. In order to simplify the 
discussion below, we consider $\hat A_j=\hbar\kappa \hat M_j$, with $j=x,y$, 
where $\kappa$ has units of wavenumber, and $\hat M_j^2=\hat 1$. 
We assume as well that the scalar potential may be considered as a 
multiple of the identity matrix (as discussed above). Removing unimportant 
global energy shifts the Hamiltonian for a free particle becomes 
\begin{equation}
\hat H=-\frac{\hbar^2}{2m}\nabla^2 \hat 1-i\frac{\hbar^2\kappa}{m}
\left ( \hat M_x \frac{\partial}{\partial x}+\hat M_y \frac{\partial}{\partial y}
\right )
\end{equation}
The atomic wavepacket can be represented by a spinorial wavefunction of the form 
\begin{equation}
\vec\Psi(\vec r,t)=\int d\vec p e^{i\vec p\cdot\vec r/\hbar}\vec\Phi(\vec p,t). 
\end{equation}
Thus, we have
\begin{equation}
\hat H\Psi(\vec r,t)=\int d\vec p \hat H_p(\vec p)\vec \Phi(\vec p,t)e^{i\vec p\cdot\vec r/\hbar},
\end{equation}
where 
\begin{equation}
\hat H_p(\vec p)\equiv \frac{p^2}{2m}\hat 1+\frac{\hbar\kappa}{m}
\left (\hat M_x p_x+\hat M_y p_y \right ).
\end{equation}
Hence for any given $\vec p$ the equation of motion 
$i\hbar\dot{\vec\Phi}(\vec p,t)=\hat H_p(\vec p) \Phi(\vec p,t)$ yields 
$\vec \Phi(\vec p,t)=\exp [i\hat H_p(\vec p)t/\hbar]\vec \Phi(\vec p,t=0)$. 
This evolution can 
be analytically obtained after diagonalizing the matrix 
$\hat H_p (\vec p)$ for every $\vec p$.

We are interested in comparing the wavepacket evolution in the presence of Abelian and
non-Abelian fields. If the fields are Abelian, i.e.\ $[\hat M_x,\hat M_y]=0$, then we may find a common eigenbasis for both 
operators, in which $\hat M_j=$diag$\{\lambda_j^+,\lambda_j^-\}$. 
As a consequence, the eigenvectors $\vec \xi_{\pm}$ of $\hat H_p(\vec p)$ are independent of $\vec p$, 
and the total wavefunction is at any time a linear combination of the form 
$\vec\Phi(\vec r,t)=\Phi_+(\vec r,t)\vec \xi_+ + \Phi_-(\vec r,t)\vec \xi_-$, where 
\begin{equation}
\Phi_\pm(\vec r,t)=e^{-i{\cal\phi}_\pm}
\int d\vec p e^{-i\frac{p^2t}{2m\hbar}}e^{i\vec p\cdot\vec r/\hbar}\Psi_{\pm}(\vec p-\vec \eta^{\pm},t=0),
\end{equation}
with $\vec \eta^{\pm}=\hbar\kappa(\lambda_x^{\pm},\lambda_y^{\pm})$, and 
${\cal\phi}_\pm=\frac{(\eta^{\pm})^2 t}{2m\hbar}+\vec\eta^{\pm}\cdot\vec r /\hbar$.
Hence, the wavepacket evolution can be considered as an independent scalar evolution for the wavepackets in each component. 
In particular, it may be easily shown that the center of mass position of the wavepacket $\Phi_\pm (\vec r,t)$ 
grows linearly in time with a velocity $(\langle \vec
p\rangle+\vec\eta^\pm)/m$. Hence the two wavepackets tend to separate  
during the time evolution.

\begin{figure}
\includegraphics[width=8 cm]{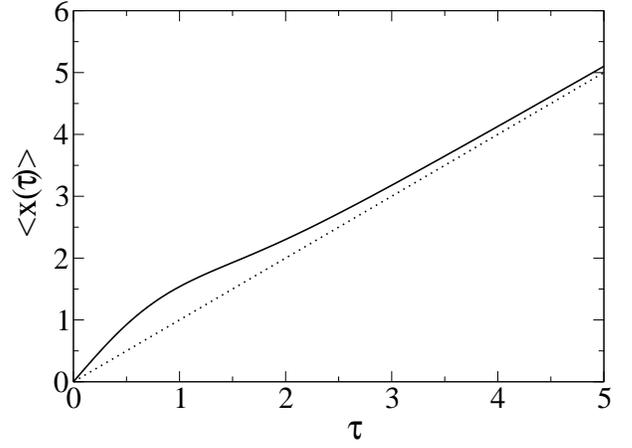}
\caption{Evolution of the center of mass coordinate $\langle x\rangle$ in units of $\sqrt{2}\Delta p/\hbar$, as a 
function of $\tau=\Delta p \kappa t/\sqrt{2}m$, for $\hat A_x=\hbar\kappa\hat\sigma_x$, and 
$\hat A_y=\hbar\kappa\hat\sigma_z$, $\eta=\pi/4$, $\varphi=0$. The dashed line is the function $f(\tau)=\tau$. 
For short times the nonlinear evolution of the center of mass becomes clear.}
\label{fig:5}
\end{figure}

The picture changes completely if $[\hat M_x,\hat M_y]\neq 0$. In this case the eigenvectors of $\hat H_p(\vec p)$ 
do depend on the momentum $\vec p$ considered, and hence the time-evolution of the wavepacket depends in a non-trivial 
way on the momentum distribution of the original wavepacket.  We analyze in particular 
the center of mass (CM) motion of the wavepacket. The $x$-coordinate of the CM after a given time $t$ 
is better calculated in the momentum representation: $\langle x\rangle_t=\langle i\hbar\partial/\partial p_x \rangle_t=
\langle e^{i\hat Ht/\hbar}
i\hbar\partial/\partial p_x
e^{-i\hat Ht/\hbar}\rangle_0$, where we have employed the Heisenberg picture. One can then easily obtain that:
\begin{equation}
\langle x\rangle_t=\langle x\rangle_0+\frac{t}{m}\langle p_x\rangle_0+
\left\langle e^{i\hat O}i\hbar\frac{\partial}{\partial p_x}
\left [e^{-i\hat O}\right ] \right\rangle_0
\end{equation} 
where $\hat O=(\kappa t/m)(\hat M_x p_x+\hat M_y p_y)$. The last term in the previous equation leads to non-trivial effects, which 
are easily illustrated by considering the particular example $\hat M_x=\hat \sigma_x$, $\hat M_y=\hat\sigma_z$:
\begin{eqnarray}
\langle x\rangle_t&=&\langle x\rangle_0+\frac{t}{m}\langle p_x\rangle_0+
\frac{\hbar\kappa t}{m} \left \{  \langle c^2\hat\sigma_x+sc\hat\sigma_z \rangle_0\right \delimiter 0 \nonumber \\
&+& \left\delimiter 0 \left\langle \frac{\sin 2q}{2q}s^2\hat\sigma_x-\frac{\sin 2q}{2q}sc\hat\sigma_z-\frac{\sin^2 q}{q}s\hat\sigma_y \right \rangle _0
\right\},
\end{eqnarray}
where $c=p_x/p$, $s=p_y/p$, $q=\kappa t p /m$, and $p^2=p_x^2+p_y^2$. Let us consider an initial Gaussian wavepacket 
$$
\vec\Psi(\vec r)=\Psi(\vec r)
\left( \begin{array}{c}
\cos\eta e^{i\varphi/2} \\
\sin\eta e^{-i\varphi/2} \end{array} \right),
$$
where  $\Psi(\vec r)$ is a Gaussian centered in $x=y=0$ and with the Fourier-Transform
$\Phi(\vec p)\sim \exp (-p^2/\Delta p^2)$. 
Then:
\begin{equation}
\langle x\rangle_\tau=\frac{\hbar}{\sqrt{2}\Delta p}\tau \left [
1+\frac{\sqrt{\pi}}{2}\frac{e^{-\tau^2}}{\tau}{\rm erfi}(\tau)
\right ]\sin 2\eta\cos\varphi,
\end{equation}
where erfi is the imaginary error function, and $\tau=\frac{\Delta p\kappa}{\sqrt{2}m}t$. 
Note, that contrary to the Abelian case, 
we have two inherently non-Abelian effects. On one hand, the evolution of the 
center of mass motion is in general a non-trivial non-linear function of time. However, 
for $\tau\gg 1$ a linear behavior $\langle x\rangle_\tau\simeq \frac{\hbar}{\sqrt{2}\Delta p}\tau$ 
is recovered, i.e.\ there is a characteristical transient stage where an inherently non-Abelian-induced 
non-linear CM evolution occurs (see Fig.~\ref{fig:5}). 
On the other hand, contrary to the Abelian (or scalar) evolution, 
the evolution of the CM motion depends on the initial width $\Delta p$ of the momentum distribution. This 
effect can be traced back to the dependence of the eigenstates $\vec\xi_\pm$ on $\vec p$.

\begin{figure}
\includegraphics[width=8 cm]{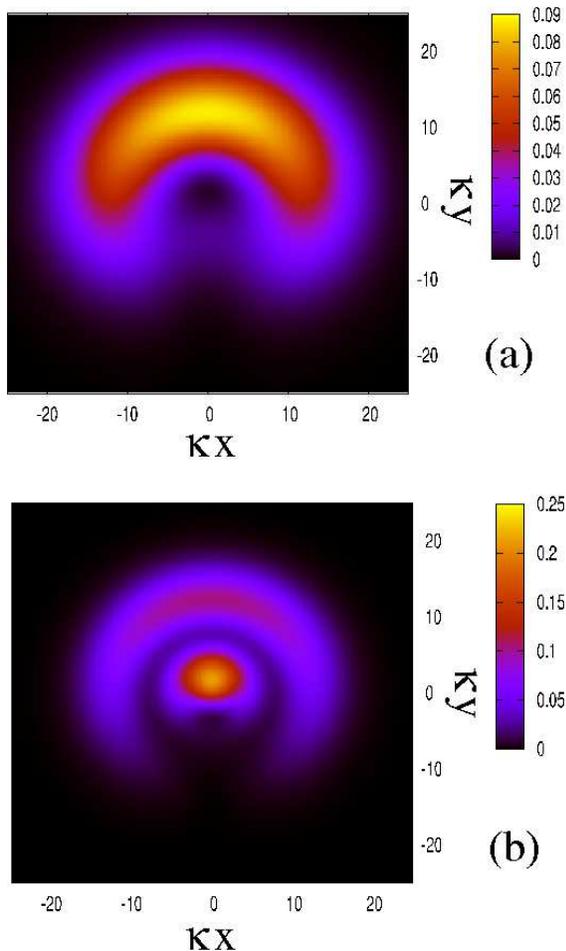}
\caption{(Color online) Total density after $t=10 (2m/\hbar\kappa^2)$, for (a) $\Delta p=0.2\hbar\kappa$ and 
(b) $\Delta p=0.6\hbar\kappa$. At $t=0$, $\eta=0$, $\varphi=0$ 
and $\langle\vec p\rangle=0$. In the strong non-Abelian case the wave packet expands asymmetrically. In an Abelian situation with a radially symmetric effective magnetic field the expansion would be symmetric.}
\label{fig:6}
\end{figure}

A third effect can be observed if we consider a Gaussian wavepacket with an initial $<p_y>_0\neq 0$. 
In that case, if $<p_x>_0=0$, one obtains:
\begin{equation}
\langle x\rangle_t=\frac{\hbar\kappa}{t} \left [ 
1-\langle \left ( 
\frac{\sin 2q}{2q}-1
\right )
\frac{p_y^2}{p^2}
\rangle_0
\right ]\sin 2\eta\cos\varphi.
\end{equation}
Hence the $x$-dynamics depends on the momentum distribution in the $y$-direction, contrary to 
the case of Abelian evolution.

Note that the details of the momentum distribution play a very important role in the 
wavepacket evolution in non-Abelian gauge fields. Obviously, if $|\langle \vec p \rangle|\gg\kappa$ 
the non-Abelian effects become negligible. But even if $|\langle \vec p
\rangle|\lesssim\kappa$, an Abelian evolution is recovered 
if $\Delta p\ll\kappa$, i.e.\ the non-Abelian effects are clearer 
for wavepackets which at $t=0$ are localized in space with uncertainties $\lesssim 1/\kappa$.
The latter effect may be explained, because 
if $\Delta p\ll\hbar\kappa$ then $\hat H_p$ may be (to a good approximation) simultaneously 
diagonalized for all relevant values of $\vec p$ in the distribution, and hence again two separated 
wavepackets as those for the Abelian evolution are recovered. In addition, it is important to realize that the particular evolution also
depends on the initial spinor configuration of the wavepacket (although this dependence is not inherently non-Abelian since it also occurs in the Abelian evolution).

Fig.~\ref{fig:6} shows the results of our numerical simulations of the wavepacket evolution 
for the Gauge field discussed above.  Note that contrary to the usual 
Abelian (or scalar) evolution, there is a stark difference in the 
evolution of the shape of the wavepacket for different values of the momentum
spreading $\Delta p/\kappa$.

The non-Abelian character of the gauge field leads also to interesting
effects in the reflection of atomic wavepackets. Ultra cold atomic wavepackets
can be reflected at laser or magnetic mirrors \cite{aminoff1993,bongs1999,arnold2006}. For typical
situations the reflection of the center of mass of the wavepacket can be
considered as specular, i.e.\ the angle of reflection of the wavepacket with the
normal vector of the mirror is exactly minus the angle of incidence of the
original wavepacket. Mathematically, the reflection can be considered as the
superposition (in absence of mirror) of the original wavepacket and an image
wavepacket travelling with opposite momentum and with a dephase $\pi$. 
For the case of wavepackets in non-Abelian gauge fields, the effect of the
mirror cannot be mimicked by this image picture (since contrary to the scalar
case, a sinusoidal solution is not an eigenstate of $\hat H_p$). As a
consequence the intuitive specular-reflection picture must be revised 
in the case of wavepackets in non-Abelian gauge fields, even for the cases 
discussed below, in which both internal components experience exactly the same 
mirror potential.  

\begin{figure}
\includegraphics[width=8.5 cm]{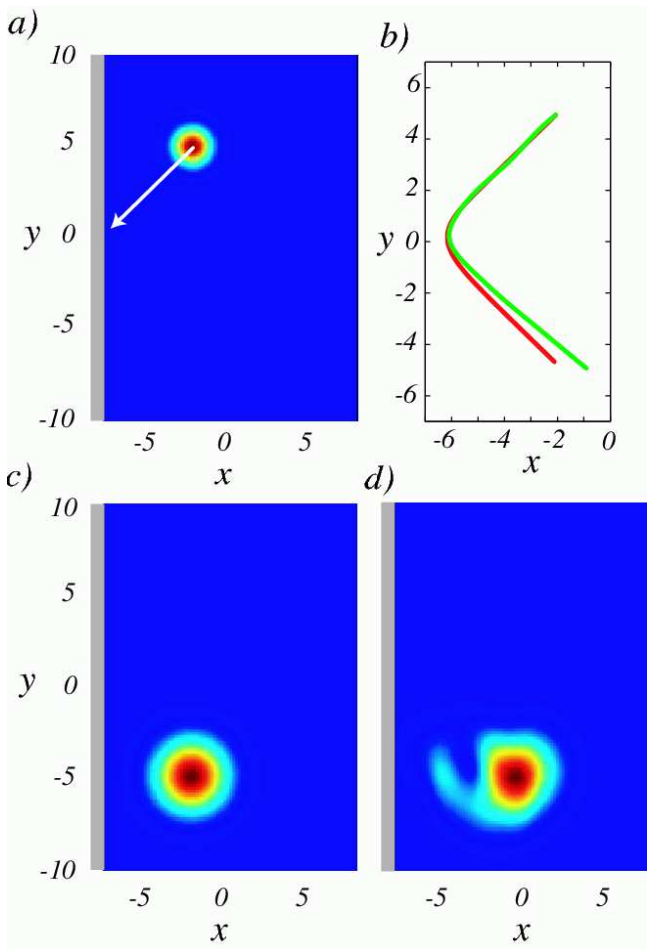}
\caption{(Color online). The reflection dynamics of a non-Abelian 
wave packet compared to a zero gauge field situation. The reflection 
takes place at $x=-7$ where a steep potential is envisaged (gray area). 
The parameters were chosen to be $\kappa=1$, $\Delta p=1$ and initial 
momentum ${\vec p_0}=-\frac{8}{\sqrt{2}}({\bf\hat x}+{\bf\hat y})$. a) The 
initial density distribution of the atomic cloud. The initial momentum 
kick is  indicated by the arrow. b) The non-Abelian path of the center
of mass, the inner (green) path, for the reflection is clearly different from the standard wave
packet reflection with $\kappa=0$ (red outer path). c) A snap shot of the wave packet 
at the time corresponding to the mirror image with respect to the 
$x$-axis. For $\kappa=0$ the reflected angle is the same as the 
incident angle. d) A snap shot of the wave packet at the same time 
as in c). For the non-Abelian case the reflection dynamics is 
highly non-trivial where the center of mass path no longer is 
described by an incident angle equal to the reflected angle. }
\label{fig:7}
\end{figure}

\begin{figure}[ht]
\includegraphics[width=8.5 cm]{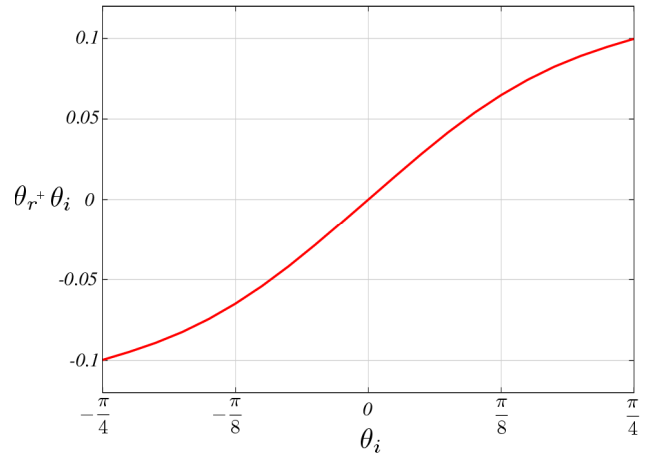}
\caption{(Color online). The reflected angle plus the incident angle, 
$\theta_r+\theta_i$, as a function of the incident angle $\theta_i$.
 The deviation from the standard case, $\theta_r+\theta_i=0$ for a 
non-Abelian system is clearly seen. The parameters were chosen to be 
$\kappa=1$, $\Delta p=1$ and initial momentum $|\vec p_0|=8$.}
\label{fig:8}
\end{figure}

Fig.~\ref{fig:7} shows the reflection of the wavepacket for $\Delta p=\hbar\kappa$
(i.e.\ for momentum spreadings for which, as discussed above, 
the non-Abelian effects are significant). It is clear from the figures that
the non-Abelian dynamics after the reflection is certainly not
trivial. Remarkably, the center-of-mass position does not show in general a
specular reflection. Fig.~\ref{fig:8} shows the sum of the angle of incidence 
and that of reflection for different incident angles in the non-Abelian
regime. For usual scalar (or Abelian) evolution, this sum equals
zero. However, 
due to inherently non-Abelian effects, this sum is significantly different
from zero. Moreover, contrary to the usual scalar (or Abelian) evolution, the 
angle of reflection crucially depends on the absolute value of the 
incoming momentum, and on the momentum spreading of the wavepacket.

\section{Conclusions}
\label{sec:5}

Summarizing, gauge fields may be generated using appropriate laser
arrangements with atoms with degenerate internal states. Using a very simple 
laser configuration, spatially homogeneous but non-Abelian vector potentials 
can be generated. In spite of this spatial homogeneity, the non-Abelian
character of the vector potentials can lead to a surprisingly rich physics for
the wavepacket dynamics of ultra cold gases. On one hand, the free expansion
dynamics of wavepackets crucially differs from what would be expected in scalar (or
Abelian) cases. In the latter, the wavepacket center-of-mass follows a linear
dependence in time. In the presence of non-Abelian fields, the wavepacket 
presents a non-linear time dependence during a transient time. In addition, and again
contrary to the scalar or Abelian case, the center-of-mass dynamics crucially
depends on the momentum spreading of the wavepacket. Moreover, in spite of the 
apparent separability of the corresponding Hamiltonian, the non-Abelian gauge
fields introduce a dependence of the dynamics in different spatial
directions. The wavepacket reflection off an
atomic mirror is also significantly distorted by the non-Abelian gauge field. 
In particular, the reflection of the center-of-mass ceases in general to be
specular, and the angle of reflection depends on the incoming velocity and 
the initial momentum spreading, which is different from the standard scalar case.
The complex interplay between external and internal dynamics should make difficult 
the read-out of non-Abelian interferometers. However, an experiment performed 
with optical tweezers, may allow for the analysis of non-Abelian effects in
the internal dynamics of the atoms. In particular, we have shown that such an
arrangement can be employed for the analysis of the equivalent of the
non-Abelian Aharanov-Bohm effect, where the final internal state of the atoms 
crucially depends on the particular path chosen.

\acknowledgements

This work was supported by the Deutsche Forschungsgemeinschaft 
(SFB-TR21, SFB407, SPP1116), the European Graduate College
``Interference and Quantum Applications'', the UK Engineering and Physical Sciences Research Council, and the Royal Society of Edinburgh.


\end{document}